\documentclass[aps,preprint,epsfig,rotate]{revtex4}
\usepackage{graphicx}
\usepackage{bm}
\usepackage{epsfig}

\begin{document}
\title{Bound state properties, positron annihilation and hyperfine structure of the four-body positronium hydrides}

\author{Alexei M. Frolov}
 \email[E--mail address: ]{alex1975frol@gmail.com} 


\affiliation{Department of Applied Mathematics \\
 University of Western Ontario, London, Ontario N6H 5B7, Canada}

\date{\today}

\begin{abstract}

Bound state properties of the ground (bound) ${}^{1}S(L = 0)-$state(s) in the four-body positronium hydrides 
${}^{1}$HPs, ${}^{2}$HPs (DPs), ${}^{3}$HPs (TPs) and MuPs are determined and investigated. By using numerical 
data from our computations of these four-body systems we have determined a number of different annihilation 
rates for each of these positronium hydrides and evaluated the hyperfine structure splitting. The properties 
of the ground (bound) states in the four-body exitonic ${}^{M}h^{+} e^{-}_{2} e^{+}$ complexes, where $M \ge 
1$ and $M \le 1$, have also been evaluated numerically. The neutral four-body systems ${}^{M}h^{+} e^{-}_{2} 
e^{+}$ with $M \gg 1$ are similar to the HPs hydrides. In particular, each of these states has only one bound 
state. We also discuss applications of the exponential and semi-exponential variational expansions 
for accurate, bound state computations of the four-body positronium hydrides. \\

\noindent
This version approximately corresponds to the final (journal) version published in: 
Journal of Molecular Physics, {\bf 120}(1), e2019337 (2022) (doi.org/10.1080/00268976.2021.2019337 ). 
In any case I recommend you to read the final/journal version of this paper. 

\end{abstract}

\maketitle

\newpage
\section{Introduction}

The positronium hydrides HPs are the Coulomb four-body systems with unit electrical charges. Each molecule 
of these positronium hydrides contains one heavy nucleus of hydrogen (either protium $p$, or deuterium $d$, 
or tritium $t$), two electrons $e^{-}$ and one positron $e^{+}$. Analogous four-leptonic system with the 
positively charged $\mu^{+}$ muon is MuPs ($\mu^{+} e^{-}_2 e^{+}$). This system is also considered here as 
a light positronium hydride. All positronium hydrides HPs and similar exitonic ${}^{M}h^{+} e^{-}_{2} e^{+}$ 
complexes are of increasing interest in solid state physics, stellar astrophysics and other areas of physics 
(see, e.g., \cite{Bhatia}, \cite{Drachman} and references therein). In solid state physics the positronium 
hydrides are important to develop and improve the general theory of four-body exitonic complexes with unit 
electrical charge which are needed to describe optical properties of many important semiconductors, e.g., 
CdS, InS, and others (see, e.g, \cite{Ziman}, \cite{Dav}). The positronium hydrides also play an important 
role in stellar astrophysics \cite{Bhatia}, \cite{Drachman}, since these four-body systems are formed in 
very large numbers in photospheres of all hot stars, e.g., in many $O-, B-$ and even in some early $A-$stars. 
By studying positron annihilation in photospheres of these stars we can investigate some crucial processes 
and physical conditions inside of such hot stars and accurately predict their time-evolution. It is 
interesting to note that annihilation of positrons in positronium hydrides in photospheres of less hot stars, 
e.g., in our Sun (spectral class G2), is extensively studied in modern papers to understand the nature and 
evaluate the rates of many fundamental processes and reactions, which proceed in outer layers of these stars 
(see, e.g., contributions and references in \cite{Bhatia}). Also this allows one to evaluate numerical 
values of some important physical parameters, e.g., the effective magnetic and electrical fields in those 
areas.     

Stability of the model positronium hydride ${}^{\infty}$HPs was shown in \cite{Ore}. Then this system has 
been investigated a number of times (see, e.g., \cite{HoDra} - \cite{Schra}). In 1992 the molecules of 
positronium ${}^{1}$HPs hydride have been created (in the collisions between positrons and methane) and 
observed in the laboratory \cite{CH4}. By replacing the methane molecule (`target' molecule) by the 
molecules of deuterium and/or tritium substituted methane, e.g., CD$_4$ and CD$_2$T$_2$, it is possible 
to create (in the same experiment) the positronium DPs and TPs hydrides. The most interesting and 
exciting property of all positronium hydrides is annihilation of the electron-positron pair(s) which is 
often called the positron annihilation. In our earlier paper \cite{OurHPs} we have considered the bound 
state properties and positron annihilation in these positronium hydrides. Later, different properties of 
the positronim hydrides have been re-evaluated in a large number of papers (see, e.g., \cite{Bhatia}, 
\cite{Bubin} - \cite{OurHPs2} and references therein). Since then the overall accuracy of our variational 
wave functions has been improved substatially. In addition to this, our current understanding of many 
processes and reactions in positronium hydrides is substantially better, than it was twenty five years ago. 
Furthermore, for the same time the masses and other physical parameters of particles, which are included in 
positronium hydrides, and numerical values of other physical constants have been changed noticeably. Now, 
by using the updated values of these constants and particle masses we want to evaluate the bound state 
properties of the HPs, DPs, TPs and MuPs hydrides to much better accuracy, than it was achieved in 
\cite{OurHPs}. For each of these positronium hydrides we also determine a few different annihilation rates 
and predict the actual hyperfine structure. We also discuss the four-body exitonic ${}^{M}h^{+} e^{-}_{2} 
e^{+}$ complexes which are similar to the regular positronium hydrides HPs (${}^{M}$H$^{+} e^{-}_{2} e^{+}$, 
or ${}^{M}$h$^{+} e^{-}_{2} e^{+}$). Here the notation H$^{+}$ and/or $h^{+}$ designates the positively 
charged hole, which has the unit electric charge, and its mass exceeds the rest mass of electron $m_e$. We 
also consider possible applications of our four-body exponential and semi-exponential variational expansions 
\cite{Our2001}, \cite{Our2003} and \cite{Fro2010} in the relative coordinates for accurate, bound state 
computations of the four-body positronium hydrides. Concluding remarks can be found in the last Section.     

\section{Bound state wave functions and properties of the positronium hydrides} 

The Hamiltonian $\hat{H}$ of the non-relativistic four-body H$^{+} e^{-}_2 e^{+}$ hydride is written 
in the form:
\begin{eqnarray}
 \hat{H} = -\frac{\hbar^2}{2 m_e} \Delta_{1} -\frac{\hbar^2}{2 m_e} \Delta_{2} -\frac{\hbar^2}{2 m_e} 
 \Delta_{3} - \frac{\hbar^2}{2 M} \Delta_{4} + \frac{e^2}{r_{12}} - \frac{e^2}{r_{13}} - 
 \frac{e^2}{r_{14}} - \frac{e^2}{r_{23}} - \frac{e^2}{r_{24}} + \frac{e^2}{r_{34}} \; , \; 
 \label{Hamilt1}
\end{eqnarray}
where $e$ is the electric charge of the positron ($-e$ is the electric charge of the electron), $m_e$ 
is the mass of the electron/positron, while $\hbar = \frac{h}{2 \pi}$ is the reduced Planck constant, 
which is also called the Dirac constant. In atomic units, where $\hbar = 1, m_e = 1$ and $e = 1$, the 
same Hamiltonian $\hat{H}$ takes the form 
\begin{eqnarray}
 \hat{H} = -\frac{1}{2} \Delta_{1} -\frac{1}{2} \Delta_{2} - \frac{1}{2} \Delta_{3} - \frac{1}{2 M} 
 \Delta_{4} + \frac{1}{r_{12}} - \frac{1}{r_{13}} - \frac{1}{r_{14}} - \frac{1}{r_{23}} - 
 \frac{1}{r_{24}} + \frac{1}{r_{34}} \; , \; \label{Hamilt}
\end{eqnarray}
where the notations (or indexes) 1 and 2 (or -) stand for the two electrons, the index 3 (or +) means the positron, 
while the index $p$ (or 4) designates the positively charged heavy nucleus of the hydrogen isotope ($p, d, t$), or 
positively charged muon $\mu^{+}$. The mass of this nucleus is denoted as $M m_e$ (or $M$ in atomic units). The same 
system of notations is used everywhere below in this paper. As mentioned above in this study we consider the bound 
state properties of the following positronium hydrides ${}^{1}$HPs, DPs, TPs and MuPs. The model four-body 
${}^{\infty}$HPs system with the infinitely heavy hydrogen nucleus is also investigated. In general, each of these 
hydrides has only one ground (bound) ${}^{1}S(L = 0)-$state (or $1^{1}S(L = 0)-$state). The first step of our 
procedure is to obtain solutions of the corresponding Schr\"{o}dinger equation $H \Psi = E \Psi$ for bound states 
($E < 0$). In fact, by solving this equation we determine the total energy $E$ and wave function $\Psi$ of the ground
(bound)${}^{1}S(L = 0)-$state for each of the postronium hydrides. At the second stage we determine various bound 
state properties of these positronium hydrides by calculating the corresponding expectation values with the wave 
functions $\Psi$ which has been derived at the first stage of our procedure. In general, the expectation value $A = 
\langle \hat{A} \rangle$ of some physical operator $\hat{A}$, which does not depend explicitly upon the time $t$, is 
defined as follows
\begin{eqnarray}
  A = \langle \hat{A} \rangle = \frac{\langle \Psi \mid \hat{A} \mid \Psi \rangle}{\langle \Psi \mid 
  \Psi \rangle} = \langle \tilde{\Psi} \mid \hat{A} \mid \tilde{\Psi} \rangle \; , \; \label{expect}
\end{eqnarray} 
where $\Psi$ is the wave function of the given (bound) state and the bound state wave function $\tilde{\Psi}$ 
has unit norm. The expectation value $A = \langle \hat{A} \rangle$ from Eq.(\ref{expect}) is often called the 
bound state property of this system in a given (bound) state. The bound state properties of the actual 
positronium hydrides HPs have been determined as described above. By choosing different quantum operators 
$\hat{X}$ in Eq.(\ref{expect}) we can evaluate numerically a large number of bound state properties, e.g., 
all inter-particle distances, interparticle delta-functions and other similar values. The bound state 
properties of the positronium hydrides are of great interest, since they allow one to describe the actual 
geometry of these four-body systems and evaluate a number of actual physical properties, including different 
annihilation rates, hyperfine structure splitting, etc. In addition to this, it is interesting to compare the 
bound state properties of different positronium hydrides and investigate their mass dependence upon variations 
of the proton mass.  

In calculations performed for this study we have applied the four-body variational expansion written in 
six-dimensional gaussoids of the relative interparticle coordinates $r_{ij} = \mid {\bf r}_i - {\bf r}_j 
\mid = r_{ji}$, where $(ij)$ = (12), (13), (23), (14), (24) and (34) and ${\bf r}_i$ are the Cartesian 
coordinates of the particle $i$, where $i$ = 1, 2, 3, 4. Note that each of the six relative coordinates 
$r_{ij}$ is translationally and rotationally invariant. This fact allows one to use the Hamiltonian 
written in the Cartesian coordinates in the form of Eqs.(\ref{Hamilt1}) and (\ref{Hamilt}). The 
variational expansion in multi-dimensional guassoids was proposed, developed and applied to various 
nuclear and atomic systems by N.N. Kolesnikov and his co-workers in the middle of 1970's (see, e.g., 
\cite{Kol1}, \cite{KT} and earlier references therein). The explicit form of this variational expansion 
in the case of four-body systems is 
\begin{eqnarray}
 \Psi = {\cal A}_S \sum^{N}_{k=1} C_k \exp( - \alpha^{(k)}_{12} r^{2}_{12} - \alpha^{(k)}_{13} r^{2}_{13} - 
 \alpha^{(k)}_{14} r^{2}_{14} - \alpha^{(k)}_{23} r^{2}_{23} -\alpha^{(k)}_{24} r^{2}_{24} -\alpha^{(k)}_{34} 
 r^{2}_{34}) \; \; \; , \label{gauss}
\end{eqnarray}
where $C_k$ are the linear variational parameters of this expansion $(k = 1, \ldots, N$), while $\alpha^{k}_{ij}$ 
are the non-linear parameters of the variational expansion, Eq.(\ref{gauss}). For four-body systems we have six 
different interparticle distances $r_{ij}$ which are explicitly shown in the variational expansion, 
Eq.(\ref{gauss}). This explains why the basis functions in Eq.(\ref{gauss}) are often called the six-dimensional 
gaussoids written in the four-body relative coordinates $r_{ij}$ \cite{five}. The four-body systems are relatively 
simple for this variational expansion, Eq.(\ref{gauss}), since in this case we do not have any unnecessary 
relative coordinate(s). In contrast with this, for the five-, six- and many body systems the number of relative 
inter-particle coordinates $r_{ij}$ (in our three-dimensional space) exceeds the total number of independent 
`radial' coordinates. In particular, for an arbitrary five-body system one finds ten interparticle coordinates in 
our three-dimensional space, while only nine of them are truly independent. For six-body systems we have three 
additional (or unnecessary) relative coordinates. In general, the appearance of unnecessary relative coordinates 
leads to additional troubles and numerical instabilities in variational calculations during optimization of the 
non-linear parameters (for more details, see, \cite{five}).  

In applications to actual four-body systems the trial wave function, Eq.(\ref{gauss}), must be symmetrized 
(or antisymmetrized) in respect to possible presence of identical particles in such systems. In each of the 
positronium hydrides we always have two identical electrons (particles 1 and 2). Therefore, the ${\cal A}_S$ 
operator in Eq.(\ref{gauss}) should include the electron-electron symmetrization operator, i.e., ${\cal A}_S 
\simeq \frac{1}{\sqrt{2}} \Bigl(1 + \hat{P}_{12}\Bigr)$ in our current notation. All our calculations in this 
study have been performed in atomic unis, where $\hbar = 1, \mid e \mid = 1$ and $m_e = 1$. The masses of 
heavy particles used in our computations are \cite{NIST}
\begin{eqnarray}
  M_{p} &=& 1836.15267343 \; \; m_e \; , \; M_{d} = 3670.48296788 \; \; m_e\; , \; \nonumber \\
  M_{t} &=& 5496.92153573 \; \; m_e \; , \; M_{\mu} = 206.7682830 \; \; m_e \; . \; \label{masses}
\end{eqnarray} 
These masses of hydrogen isotopes are currently recommended for scientific use by CODATA/NIST in 2021/2022 
\cite{NIST}.  

In general, the overall efficiency of the variational expansion, Eq.(\ref{gauss}), substantially depends upon 
algorithms which are used to optimize the non-linear parameters in Eq.(\ref{gauss}). Recently, for four-body 
systems we have developed a number of algorithms which were found to be quite effective and fast in 
applications to various systems. The accuracy of the constructed wave functions is usually high (and even very 
high) for the total energies. The expectation values of various geometrical and dynamical properties are also 
determined (with the same wave functions) to relatively high numerical accuracy. However, some troubles can 
still be found in computations of the expectation values of some delta-functions, interparticle cusp values 
\cite{Kato} and for other similar properties (see discussion in Section VI).   

Results of our numerical computations performed for the ground ${}^{1}S(L = 0)-$states in the positronium 
hydrides are shown in Table I. These results include the total energies and other bound state properties of 
all positronium hydrides discussed in this study. As mentioned above in these hydrides the index 4 designates 
the heavy, positively charged nucleus of the hydrogen isotope, i.e., protium $p$, deuterium $d$ and tritium 
$t$. We also determine the bound state properties of the model ${}^{\infty}$H$^{+} e^{-}_2 e^{+}$ hydride 
and four-leptonic MuPs hydride which is the $\mu^{+} e^{-}_2 e^{+}$ ion. In the last two cases in Table I 
the index $p$ (or 4) stands for the infinitely heavy ${}^{\infty}$H$^{+}$ nucleus (model nucleus) and for 
the positively charged muon $\mu^{+}$, respectively. The properties shown in Table I include the $\langle 
r^{k}_{ij} \rangle$ expectation values, where $k$ is the power of the interparticle $(ij)$-distances. In 
calculations performed for this study we used $k$ = -2, -1, 1, 2, 3 and 4. In Table I one also finds the 
expectation values of single particle kinetic energies $\langle - \frac12 \nabla^{2}_i \rangle$ determined 
for the both electron and positron. Table II contains the expectation values of of all interparticle 
delta-functions $\langle \delta_{ij} \rangle$ and triple electron-positron delta-functions $\langle 
\delta_{+--} \rangle$. These delta-functions are needed in the following Sections to determine the 
corresponding annihilation rates. Physical meaning of all bound state properties shown in Table I is 
clear from the notation used. Analytical formulas for all matrix elements used in computations of these 
bound state properties can be found in \cite{OurHPs}. Note that the two-particle cusp values \cite{Kato} 
cannot be computed directly with the use of variational expansion, Eq.(\ref{gauss}), since the 
corresponding expectation values equal zero identically. This is an obvious disadvantage of the 
variational expansion, Eq.(\ref{gauss}), in applications to the few- and many-body Coulomb systems. 

However, this disadvantage is not critical and all similar `problematic' expectation values can be 
evaluated indirectly by using one of the two following procedures: (a) re-expansion of the wave 
functions which is already represented in terms of variational expansion, Eq.(\ref{gauss}), and (b) 
calculations of the expectation values of some other operators and their commutators. Then, at the 
second step of the procedure (b) the expectation values, which are needed for our evaluations, are 
constructed as linear combinations of the expectation values of these operators. In particular, for 
the interparticle delta-functions in Coulomb few-body systems such an approach has been developed 
and tested by R.J. Drachman and his colleagues (see, e.g., \cite{DraSu}).  

\section{Positron annihilation}

The most interesting property of each positronium hydride is annihilation of the electron-positron pair, 
which is often called the positron annihilation, for short. In general, annihilation of the 
electron-positron pair(s) in positronium hydrides proceeds mainly with the emission of two high energy 
photons, or two annihilation $\gamma-$quanta. In this study we restrict ourselves to the analysis of 
positron annihilation from the bound states of positronium hydrides only. An alternative process which
is called annihilation-in-flight is not discussed here (see, e.g., \cite{Fer} and \cite{Bha16}). Similar
process proceeds, e.g., during collisions between negatively charged hydrogen ion and fast positrons. 
The probability of the two-photon annihilation per second is called the two-photon annihilation rate 
$\Gamma_{2 \gamma}$. Ii is clear that the $\Gamma_{2 \gamma}$ value is a unique property of any given 
bound state. For the ground (bound) ${}^{1}S(L = 0)$ state in the positronium hydrides the explicit 
formula for the $\Gamma_{2 \gamma}$ rate can be written in the following form (see, e.g., \cite{Heitl} 
- \cite{Grein} and \cite{BLP})
\begin{eqnarray}
 \Gamma_{2 \gamma} = n \; m \; \pi \; \alpha^{4} c a^{-1}_{0} \langle \delta({\bf r}_{+-}) \rangle = 
 2 \pi \alpha^{4} c a^{-1}_{0} \langle \delta_{+-} \rangle \; \; , \; \label{twopho}
\end{eqnarray} 
which includes the expectation value of the electron-positron delta-function $\langle \delta({\bf 
r}_{+-}) \rangle = \langle \delta_{+-} \rangle$ expressed in atomic units. This expectation value 
must be evaluated numerically for the ground ${}^{1}S(L = 0)$ state in the HPs hydride. In this 
formula the notation $\alpha = 7.2973525693 \cdot 10^{-3}$ stands for the dimensionless 
fine-structure constant, $c = 2.99792458 \cdot 10^{10}$ $cm \cdot sec^{-1}$ is the speed of light 
in vacuum and $a_{0} = 5.29177210903 \cdot 10^{-9}$ $cm$ is the Bohr radius \cite{NIST}. Here and 
everywhere below in this study, all numerical values for physical constants have been taken from 
recent set of physical constants values published by NIST \cite{NIST}. Note that these values are 
currently recommended for scientific use by CODATA/NIST in 2021/2022 \cite{NIST}. 

Also, in Eq.(\ref{twopho}) and in all formulas below the notation $n$ stands for the total number of 
bound electrons in atomic system, while $m$ denotes the total number of positrons bound in the same 
atomic system. The product $m n$ is the total number of electron-positron pairs in this atomic 
system. In fact, in all positronium hydrides HPs we have $n = 2$ and $m = 1$, i.e., the product 
$m n$ equals two. The notation $\Gamma_{2 \gamma}$ is very convenient and everywhere below in this 
study we shall apply the notation $\Gamma_{k \gamma}$ to designate the $k-$photon ($k \ge 0$) 
annihilation rate averaged over orientations of the spins of all electrons and positrons 
participating in the process. Analogous formula for the three-photon annihilation rate $\Gamma_{3 
\gamma}$ in the posironium hydrides is 
\begin{eqnarray}
 \Gamma_{3 \gamma} = n m \; \frac{4 (\pi^{2} - 9)}{3 \pi} \pi \alpha^{5} c a^{-1}_{0} \langle 
 \delta({\bf r}_{+-}) \rangle = \frac{8 (\pi^{2} -  9)}{3} \alpha^{5} c a^{-1}_{0} \langle 
 \delta_{+-} \rangle \; \; . \; \label{threpho}
\end{eqnarray} 

These two largest annihilation rates $\Gamma_{2 \gamma}$ and $\Gamma_{3 \gamma}$ are crucially 
important to describe positron annihilation in any atomic and/or molecular system. Furthermore, 
if we know these two annihilation rates, then we can also evaluate the four- and five-photon 
annihilation rates, i.e., the $\Gamma_{4 \gamma}$ and $\Gamma_{5 \gamma}$ values, respectively. 
The corresponding approximate formulas for these values have been derived in \cite{1983} 
\begin{eqnarray}
  \Gamma_{4 \gamma} = 0.274 \Bigl(\frac{\alpha}{\pi}\Bigr)^{2} \Gamma_{2 \gamma} \; \; \; 
  {\rm and} \; \; \;  
  \Gamma_{5 \gamma} = 0.177 \Bigl(\frac{\alpha}{\pi}\Bigr)^{2} \Gamma_{3 \gamma} \; \; . 
  \; \label{4and5pho}
\end{eqnarray} 
The formulas Eqs.(\ref{threpho}) and (\ref{4and5pho}) are used in this study to determine the 
$\Gamma_{3 \gamma}, \Gamma_{4 \gamma}$ and $\Gamma_{5 \gamma}$ annihilation rates for all 
positronium hydrides mentioned in this study. 

Currently, the formula, Eq.(\ref{twopho}), for the two-photon annihilation rate is used only 
occasionally. Instead, the following formula for the $\Gamma_{2 \gamma}$ annihilation rate (in 
the Ps$^{-}$, Ps$_2$ and HPs systems) is applied in modern calculations and experimental 
measurements
\begin{eqnarray}
 \Gamma_{2 \gamma} = n \; m \; \pi \alpha^{4} c a^{-1}_{0} \Bigl[ 1 - \frac{\alpha}{\pi} \Bigl( 
  5 - \frac{\pi^{2}}{4} \Bigr) \Bigr] \; \langle \delta_{+-} \rangle \; \; . \; \label{gamma2}
\end{eqnarray} 
This formula contains the two-photon annihilation rate, Eq.(\ref{twopho}), and also includes the 
lowest order QED correction to that value \cite{HaBr}. In our calculations of $\Gamma_{2 \gamma}$ 
performed for this study we also apply the formula, Eq.(\ref{gamma2}). For many actual systems it 
is also interesting to evaluate the total annihilation rate $\Gamma$, which is the sum of all 
partial annihilation rates. In reality, the four-, five- and other multi-photon annihilation rates 
are significantly smaller than the sum of the two- and three-photon annihilation rates. This allows 
one to write the following approximate expression for the total annihilation rate of the 
electron-positron pair in arbitrary atom and/or molecule 
\begin{eqnarray}
 \Gamma \approx \Gamma_{2 \gamma} + \Gamma_{3 \gamma} = n \; m \; \pi \alpha^{4} c a^{-1}_{0} \Bigl[ 
 1 - \alpha \Bigl( \frac{17}{\pi} - \frac{19 \pi}{12} \Bigr) \Bigr] \; \langle \delta_{+-} \rangle \; 
 \; . \; \label{gammatot}
\end{eqnarray} 
This formula has a very good numerical accuracy for arbitrary $n$ and $m$, and it is often used in 
various applications to all positron containing few-body systems \cite{Fro2009}. The total 
annihilation rate $\Gamma$ was measured in early 1980's \cite{Mils1}, \cite{Mils2} for the three-body 
Ps$^{-}$ ion. In regular experiments and observations with poly-electrons, atoms and molecules, which 
contain positron(s), one usually deals with the total annihilation rate $\Gamma$. This includes the 
three-body Ps$^{-}$ ion ($m n$ = 2), four-body bi-positronium molecule Ps$_2$ ($m n$ = 4), positronium 
hydrides HPs ($m n$ = 2) and other positron-containing atoms and molecules which are known in 
astrophysics and solid state physics. All ahhihilation rates mentioned here and below (the both 
$\Gamma^{a}_{1 \gamma}$ and $\gamma^{F}_{1 \gamma}$ rates) can be found in Table III. 

\subsection{One- and zero-photon annihilation rates}

As mentioned above all positronium hydrides are the bound Coulomb four-body systems which contain three 
light particles and one heavy (or central), positively charged particle. In each of these hydrides all 
four particles interact with each other and, in general, such interactions are not small. This means 
that annihilation of the electron-positron pair in each of these positronium hydrides may also proceed 
with the emission of either a single photon (one-photon annihilation), or even with no emission of 
photon(s) et al. Briely, zero-photon annihilation in postronium hydrides HPs means that the both photons 
emitted during two-photon annihilation are absorbed by the remaining particles: one by the electron and 
another by the heavy nucleus. Let us briefly discuss annihilation of the electron-positron pair in the 
HPs hydrides which proceeds with no emission of one and zero photon(s). The corresponding annihilation  
rates are designated below as $\Gamma_{1 \gamma}$ (with some additional indexes) and $\Gamma_{0 \gamma}$,  
respectively. It is easy to predict that the numerical values $\Gamma_{1 \gamma}$ and $\Gamma_{0 \gamma}$  
are very small. Nevertheless, in a number of problems and processes known in astrophysics, solid state 
physics, etc, these one- and zero-photon annihilations play some roles. For the first time, the one-photon 
annihilation rate $\Gamma_{1 \gamma}$ in the three-body Ps$^{-}$ ion was determined in \cite{Kru}, 
\cite{Fro1995} (see, also \cite{Fro1999})$^{1}$. In earlier paper \cite{Chu} the authors took 
into account twice less Feynman diagrams than in \cite{Kru} and obtained a different result. 

The same one-photon annihilation is possible in the four-body positronium hydrides and the corresponding 
annihilation rate $\Gamma^{a}_{1 \gamma}$ was evaluated in our earlier paper \cite{OurHPs}. In this case 
the electron-positron pair annihilates in the nearest vicinity of another (second) electron in the HPs 
hydride. The value of $\Gamma^{a}_{1 \gamma}$ is given by the formula \cite{Kru} which is written in the 
following form (in atomic units) \cite{Fro1995}    
\begin{eqnarray}
 \Gamma^{a}_{1 \gamma} = \frac{64 \pi^2}{27} \; \alpha^{8} c a^{-1}_{0} \langle \delta_{+--} \rangle 
 \approx 1065.756921 \cdot \langle \delta_{+--} \rangle \; \; , \; \label{gamma1a}
\end{eqnarray} 
where the notation $\langle \delta_{+--} \rangle$ is the expectation value of the three-particle $e^{-} 
e^{+} e^{-}$ delta-function which must be expressed in atomic units. The formula, Eq.(\ref{gamma1a}), 
is exactly the same as it was derived for the three-body Ps$^{-}$ ion. This triple delta-function is 
determined for the ground (bound) ${}^{1}S(L = 0)-$state of the positronium hydride(s). In general, the 
knowledge of the $\langle \delta_{+--} \rangle$ expectation value allows one to evaluate the one-photon 
annihilation rate $\Gamma^{(I)}_{1 \gamma}$ in the three-body Ps$^{-}$ ion, four-body bi-positronium 
molecule Ps$_{2}$, all four-body positronium hydrides HPs and other few-body compounds which contain 
positron(s) and more than one bound electron.   

In all positronium hydrides mentioned above another one-photon annihilation of the electron-positron 
$(e^{-}, e^{+})$-pair is also possible. Indeed, let us assume that positron approaches the electron which 
is bound by the Coulomb field of the heavy, positively charged atomic nucleus. Therefore, the probability 
(or rate) of such an annihilation can be determined exactly in the same way as we calculate the cross 
section(s) of atomic photoionization. During such computations the positron must be considered as an 
electron which moves backwards in time. This one-photon annihilation was studied and described in detail 
in \cite{Fermi}, \cite{Beth2} (see, also \cite{Heitl} and references therein). For few-electron 
quasi-atomic systems this one-photon annihilation rate, which is often called the Fermi annihilation 
rate, is written in the form (in atomic units, see, e.g., \cite{Heitl})
\begin{eqnarray}
 \gamma^{F}_{1 \gamma} = \frac{4 \pi}{3} \; n \; m \; \alpha^{8} Q^{5} \delta_{+-} v_{+-} 
\Bigl(\frac{c}{a_0}\Bigr) \; , \; \label{gamma1b}
\end{eqnarray} 
where the notation $n$ stands for the total number of bound electrons, while $m$ denotes the total number 
of bound positrons in this atom/molecule, while $Q$ is the electric charge of the heavy, central nucleus. 
Also, in this equation $\delta_{+-}$ is the electron-positron delta-function and $v_{+-}$ is the relative 
velocity of the two particles (electron and positron) at the point of electron-positron coalesence. The 
exact formula for the $\Gamma^{F}_{1 \gamma}$ rate takes the form 
\begin{eqnarray}
 \Gamma^{F}_{1 \gamma} = \frac{4 \pi}{3} \; n \; m \; \alpha^{8} Q^{5} \; \langle \delta_{+-} 
 \frac{p_{+-}}{m_e} \rangle \; \Bigl(\frac{c}{a_0}\Bigr) = \frac{4 \pi}{3} n m \alpha^{8} Q^{5} \mid 
 \nu_{+-} \; \mid \langle \delta_{+-} \rangle \; \Bigl(\frac{c}{a_0}\Bigr) \; , \; \label{Gamma1b}
\end{eqnarray} 
where $\nu_{+-} = \frac{\langle \delta_{+-} \nabla_{+-} \rangle}{\langle \delta_{+-} \rangle}$ is the 
electron-positron cusp (its definition can be found, e.g., in \cite{Kato} and \cite{Fro1999}). For highly 
accurate, bound state wave functions the electron-postiron cusp always equals $-\frac12 = - 0.5$ (to very 
high accuracy). In other words, it almost coincides with the analytically predicted cusp value between 
two point, electrically charged particles $i$ and $j$ which equals $\nu_{ij} = q_i q_j \frac{m_i m_j}{m_i 
+ m_j}$ \cite{Kato}. For the positronium hydrides, where $Q = 1, n = 2$ and $m = 1$, the last formula 
takes the form 
\begin{eqnarray}
 \Gamma^{F}_{1 \gamma} = \frac{2 \pi}{3} \; n \; m \; \alpha^{8} \; \langle \delta_{+-} \rangle \; 
 \Bigl(\frac{c}{a_0}\Bigr) = \frac{4 \pi}{3} \alpha^{8} \; \langle \delta_{+-} \rangle \; 
 \Bigl(\frac{c}{a_0}\Bigr) \; , \; \label{Gamma1bf}
\end{eqnarray} 
where all expectation values are determined for the ground (bound) state in each of the positronium hydrides. 
Note that this one-photon annihilation rate is proportional to the expectation value of the electron-positron 
delta-function $\langle \delta_{+-} \rangle$, while the $\Gamma^{a}_{1 \gamma}$ annihilation rate, 
Eq.(\ref{gamma1a}), is proportional to the expectation value of the triple electron-positron-electron 
delta-function $\langle \delta_{+--} \rangle$. Such a difference in formulas follows from the fundamental 
difference between these two annihilation processes. The one-photon (Fermi) annihilation rate $\Gamma^{F}_{1 
\gamma}$ has never been evaluated for the positronium hydrides HPs. In the light positronium systems, e.g., 
in the Ps$^{-}$ ion and/or in the Ps$_{2}$ molecule, the one-photon Fermi annihilation is impossible.  

In our previous paper \cite{OurHPs} we also evaluated another one-photon annihilation rate $\Gamma^{(b)}_{1 
\gamma}$ and zero-photon annihilation rate $\Gamma^{(b)}_{0 \gamma}$. These evaluations were essentially 
based on our earlier results obtained for the four-body, bi-positronium Ps$_2$ molecule \cite{Fro1995}. Later, 
we have found that `analogies' between the light Ps$_2$ system and heavy (one-center) positronium hydrides HPs 
do not work well for one-center HPs systems, and this may lead to very large errors in similar evaluations. In 
particular, in the HPs hydrides the Coulomb field of the central (very heavy) hydrogen nucleus always play a 
substantial role in these annihilation processes (in contrast with the Ps$_2$ system). Currently, we derive 
more accurate formulas which can be used to describe these annihilation rates in HPs hydrides. The new 
analysis of these annihilation processes in the HPs hydrides and calculations of their annihilation rates will 
be discussed elsewhere. 

\section{Hyperfine structures of the four-body positronium hydrides}

As mentioned above all positromium hydrides are the Coulomb four-body systems of actual, interacting particles 
and each of these particles has non-zero spin. In fact, all particles which form these positronium hydrides 
${}^{1}$HPs, ${}^{2}$HPs (DPs), ${}^{3}$HPs (TPs) and MuPs, i.e., electrons, positron, $\mu^{+}-$muonium, 
protium and tritium nuclei (but deuterium nucleus!) have spin which equals $\frac12$. The spin of the deuterium 
nucleus equals unity. Therefore, even for the bound $1{}^{1}S(L = 0)-$state in each of these positronium 
hydrides one can observe the direct spin-spin interactions between four particles in the HPs hydrides. In 
general, for the bound $S(L = 0)-$states such a spin-spin (positron-nucleus) interaction (see below) proceeds 
at very short distances between two particles with non-zero spin. Also, it is crucial to note that in all 
positronium hydrides ${}^{1}$HPs, DPs, TPs and MuPs the two bound electrons can only be in the singlet 
${}^{1}S(L = 0)-$state. Otherwise, i.e., for the triplet ${}^{3}S(L = 0)$ electron pair, the whole four-body 
HPs hydride (or H$^{+} e^{-}_2 e^{+}$ system) is not bound \cite{OurHPs}. Therefore, in each of the positronium 
hydrides there is a spin-spin interaction between positively charged positron and heavy nucleus of hydrogen 
isotope (or $\mu^{+}$-muon). In general, such an interaction is relatively small $\simeq \alpha^{2}$ and it is 
proportional to the expectation value of the positron-nucleus delta-function. This follows from the explicit 
form of the Hamiltonian $H_{HF}$, which describes the hyperfine structure of the positronium hydrides. In 
atomic units this Hamiltonain $H_{HF}$ is 
\begin{eqnarray}
 H_{HF} &=& - \sum_{(ij)} a_{ij} \; ({\bf s}_{i} \cdot {\bf s}_{j}) = - a \; ({\bf s}_{+} \cdot {\bf I}_H) - 
 b \; ({\bf s}_{+} \cdot {\bf S}_{-}) - c \; ({\bf S}_{-} \cdot {\bf I}_H) = - a \; ({\bf s}_{+} \cdot {\bf 
 I}_H) \; \; \nonumber \\ 
 &=& - \frac{8 \pi \alpha^{2}}{3} \mu^{2}_{B} \; \Bigl(\frac{g_{+}}{m_e}\Bigr) \Bigl(\frac{g_{H}}{M_p}\Bigr) \; 
 \langle \delta_{+ H} \rangle \; ({\bf s}_{+} \cdot {\bf I}_H) \; \; , \; \label{hypfspl} 
\end{eqnarray} 
where $\alpha$ is the fine-structure constant, $\mu_B$ is the Bohr magneton, $M_p$ is the proton mass (at rest), 
while ${\bf S}_{-}$ is the total electron spin, which equals zero for the singlet electron-electron pair (for 
the triplet electronic states this formula, Eq.(\ref{hypfspl}), has a different form). In atomic units the 
absolute value of the Bohr magneton $\mu_{B} = \frac{e \hbar}{2 m_e}$ equals $\frac12$ (exactly). The notation 
$\langle \delta_{+ H} \rangle$ stands for the expectation value of the positron-nucleus delta-function 
expressed in atomic units. This $\delta-$function must be determined for the ground $1^{1}S(L = 0)-$state of 
the HPs hydride. The notations $g_{+}$ and $g_{H}$ stand for the gyromagnetic ratios of the positron and heavy 
nucleus of the hydrogen isotope, $M_p$ is the proton mass (in atomic units), $m_e$ is the positron mass and 
$I_H$ is the nuclear spin of the hydrogen isotope which is integer (or semi-integer) positive value. The 
nuclear $g-$factor (or $g_H$ factor in Eq.(\ref{hypfspl})) is the ratio of magnetic moment of this nucleus (in 
nuclear magnetons) to its maximal spin value $I_N (= I_H)$ \cite{Fro2012}. In positronium hydrides we have 
$I_H = \frac12, I_T = \frac12$ and $I_D$ = 1, while in the positronium MuPs hydride the factor 
$\Bigl(\frac{g_{H}}{M_p}\Bigr)$ must be replaced by the analogous muonic factor 
$\Bigl(\frac{g_{\mu}}{m_{\mu}}\Bigr)$. 

The hyperfine structure splitting $\Delta E_{HF}$ in the HPs hydrides is the difference between two bound states 
with the total spin $I_H + \frac12$ and $I_H - \frac12$, respectively. This difference is 
\begin{eqnarray}
 \Delta E_{HF} = \frac{2 \pi \alpha^{2}}{3} \; \Bigl(\frac{g_{+}}{m_e}\Bigr) \Bigl(\frac{g_{H}}{M_p}\Bigr) \; 
 F_{Ry} \; \langle \delta_{+ H} \rangle \; ({\bf s}_{+} \cdot {\bf I}_H) \; \; MHz = {\cal A} \; \langle 
 \delta_{+ H} \rangle \; ({\bf s}_{+} \cdot {\bf I}_H) \; \; MHz \; , \; \label{hypfspld}
\end{eqnarray} 
where the constant ${\cal A}$ is 
\begin{eqnarray}
 {\cal A} = \frac{2 \pi \alpha^{2}}{3} \; \Bigl(\frac{g_{+}}{m_e}\Bigr) \Bigl(\frac{g_{H}}{M_p}\Bigr) \; F_{Ry} 
 \nonumber 
\end{eqnarray} 
Numerical values of the $g-$factors in this formula are \cite{NIST}
\begin{eqnarray}
  g_{e} &=& -2.00231930436256 \; \; , \; \; g_{\mu} = -2.0023318418 \; \; , \; \; g_{p} = 5.5856946893 \; 
 \nonumber \\
 g_{d} &=& 0.8574382338 \; \; , \; \; g_{t} = 5.957924931 \; \; , \; \label{g-fact}
\end{eqnarray} 
while the numerical values of the fine-structure constant $\alpha$ and proton mass $M_p$ have been taken from 
the previous Sections. The conversion factor (or Rydberg factor) is $F_{Ry} = 2 Ry \cdot c = 2 \times 
3.2898419602508 \cdot 10^{15} \; Hz \; = 6.5796839205016 \cdot 10^{9}$ $MHz$ \cite{NIST} in Eq.(\ref{hypfspld}) 
allows one to re-calculate the hyperfine structure splitting from atomic units to $MHz$ (or $kHz$) which are 
the traditional units for these values. The factor $({\bf s}_{+} \cdot {\bf I}_H)$ in Eq.(\ref{hypfspld}) is 
the scalar product of the vector of positron spin ${\bf s}_{+}$ and nuclear spin ${\bf I}_H$. This factor 
equals $({\bf s}_{+} \cdot {\bf I}_H) = s_{+} ( 2 I_{H} + 1 ) = 1$, if $I_{H} = \frac12$, and $\frac32$, if 
$I_{H} = 1$.   

The formula, Eq.(\ref{hypfspld}), is used in this study to determine the hyperfine structure splittings in all 
positronium hydrides mentioned above. Note that for all positronium hydrides we have splitting between the two 
groups of energy levels, e.g., the upper and lower energy levels. Indeed, if $I_H$ is the spin of the nucleus, 
then one finds $I_H = \frac12$ for the $p^{+}$ and $t^{+}$ nuclei and $\mu^{+}$ muon. For the deuterium nucleus 
we have $I_D$ = 1. The positron's spin equals $\frac12$. Therefore, in the ${}^{1}$HPs, TPs and MuPs hydrides 
all states from first group have the total spin equals unity (triplet states), while the state(s) from the 
second group have the total spin equals zero (singlet states). For the deuterium DPs hydride all states from 
the first group have spin equals $\frac32$ (quartet states), while analogous states from the second group have 
spin equals $\frac12$ (doublet states). Numerical values of the hyperfine structure splitting determined for 
each of the positronium hydrides, considered in this study, can be found in Table \ref{TablHF} which also 
contains the factors ${\cal A}$ used in the formula, Eq.(\ref{hypfspld}), for each of these hydrides. 

\section{Bound states in the four-body exitonic complexes}

The neutral four-body complexes (or clusters) with unit electrical charges $A^{+} e^{-}_2 e^{+}$ are similar 
to the HPs hydrides, and they are of interest in various applications closely related to the physics and 
spectroscopy of those semiconductors, which have large and very large electric permittivity. It is shown in 
numerous papers and books on solid states physics (see, \cite{Ziman}, \cite{Dav}) that many unique optical 
properties of semiconductors can only be explained, if we allow formation of some few-body exotic systems, 
which include 'effective' electrons $e^{-}$, positively charged hole(s) $h^{+}$ and atom(s) $A$ (or ion(s) 
$A^{+}$) which represent the impurities and/or lattice defects. A number of experimental results obtained in 
solid state physics of semiconductors were explained in the late 1930's with the help of two-body exitons of 
Wannier-Mott \cite{Wan}, \cite{Mott}. These two-body exitons are similar to the electron-positron pairs in 
QED (see, e.g., \cite{Dav}). In \cite{Lamp} Lampert suggested that exitonic complexes may also include more 
than two light particles, e.g., they can include a few light particles (holes and electrons) and one heavy 
particle. Furthermore, such few-body exitonic complexes can be either mobile (light), or heavy (immovable). 
The mobile exitonic complexes usually consist of a few electrons $e^{-}$ and some light, positively charged 
hole(s) $h^{+}$. Immovable exitonic complexes also include at least one heavy, positively charged central 
atom/ion which is usually associated with either impurity, or lattice defect.

Let us consider the four-body exitonic complex $A^{+} e^{-}_2 e^{+}$ (or $h^{+} e^{-}_2 e^{+}$) which are 
similar to the positronium hydrides HPs. First, we discuss the four-body exitonic systems which contains 
two electrons $e^{-}$, one positron $e^{+}$ and one heavy, positively charged `quasi-nucleus' $A^{+}$ (or 
$h^{+}$). All these particles have unit electrical charges. The Hamiltonians of similar four-body HPs-like 
exitonic complexes are written in the following general form 
\begin{eqnarray}
  H = -\frac{\hbar^2}{2 m_e} \Bigl[\nabla^2_1 + \nabla^2_2 + \nabla^2_3 + \Bigl(\frac{m_e}{M}\Bigr) 
 \nabla^2_4 \Bigr] - \frac{e^2}{r_{41}} -  \frac{e^2}{r_{42}} - \frac{e^2}{r_{31}} - \frac{e^2}{r_{32}} 
 + \frac{e^2}{r_{21}} + \frac{e^2}{r_{43}} \; , \; \label{HamilM}
\end{eqnarray}
where $\hbar = \frac{h}{2 \pi}$ is the reduced Planck constant (or Dirac constant), $m_e$ is the electron 
mass at rest and $e$ is the electric charge of the electron. In this equation we apply the same notations 
which were used in Section II, i.e., the subscripts 1 and 2 designate the two bound electrons $e^-$, the 
subscript 3 denotes the positron $e^{+}$, while the subscript 4 means the positively charged `heavy' hole 
which has the mass $M (\ge 1)$. The electric charge of this hole equals to the electric charge of positron 
$e^{+}$. All other notations are exactly the same as they were used in Section II above. In atomic units 
(where $\hbar = 1, \mid e \mid = 1, m_e = 1$) this Hamiltonian takes the form 
\begin{eqnarray}
 H = -\frac12 \Bigl[\nabla^2_1 + \nabla^2_2 + \nabla^2_3 + \frac{1}{M} \nabla^2_4 \Bigr] - \frac{1}{r_{41}} 
 - \frac{1}{r_{42}} - \frac{1}{r_{31}} - \frac{1}{r_{32}} + \frac{1}{r_{21}} + \frac{1}{r_{43}} \; \; . \; 
 \label{HamilM1}
\end{eqnarray}
As follows from this expression the Hamiltonian operator is a regular function of the dimensionless parameter 
$\frac{m_e}{M} = \frac{1}{M}$, where the mass $M$ is expressed in the electron mass $m_e$.    

First, we investigate transformation of the light (i.e., mobile) four-body exitonic complexes into heavy exitonic 
complexes which are less mobile. For now, we shall assume that the mass of heavy particle (or the effective 
mass of the impurity) in the ${}^{M}h^{+} e^{-}_2 e^{+}$ systems varies (increases) from the unity $M_h = m_e$ 
(bi-positronium Ps$_2$) to relatively large mass values, e.g., up to $M$ = 12 $m_e$ and larger masses. It is 
interesting to investigate quantitative changes in the total energies and other bound state properties of the 
${}^{M}h^{+} e^{-}_2 e^{+}$ systems, when such a mass increases. It is clear that when the mass of heavy 
particle ($M$) increases to very large values, then the bound states properties, including the total energies, 
become similar to the analogous properties of the heavy HPs hydrides and MuPs system. However, the mass region 
$m_e \le M \le 12$ $m_e$ is critical to understand the effect of rapid changes which occur in many bound state 
properties of the ${}^{M}h^{+} e^{-}_2 e^{+}$ systems, when the heavy mass $M$ increases. Results of our 
computations performed for a number of ${}^{M}h^{+} e^{-}_2 e^{+}$ systems can be found in Table \ref{HmPs}. 
For each system shown in this Table the corresponding  results include the total energy $E$, $\langle r_{+-} 
\rangle , \langle r_{--} \rangle$ and $\langle r_{h -} \rangle$ distances and the expectation value of the 
electron-positron delta-function $\langle \delta({\bf r}_{+-}) \rangle$.  

Our results obtained for the exitonic ${}^{M}h^{+} e^{-}_2 e^{+}$ complexes can be represented 
in the form of mass-interpolation formula which provides a very good accuracy in actual 
applications. This formula has been derived to approximate the mass-dependencies of actual 
properties of the four-body ${}^{M}h^{+} e^{-}_2 e^{+}$ systems is an analytical function of 
the following dimensionless parameter $Z = 1 - \frac{2 m_e}{M}$, which is the inverse mass of 
the heaviest particle. The actual mass-dependence of the total energies of these exitonic 
complexes is accurately represented by the following mass-interpolation formula
\begin{eqnarray}
 E(Z) = \Bigl(\frac{1 + Z}{2}\Bigr) E(1) + \Bigl(\frac{1 - Z}{2}\Bigr) E(-1) + \Bigl(\frac{1 
 - Z^{2}}{4}\Bigr) \Bigl[ C_0 &+& C_1 P_{1}(Z) + C_2 P_{2}(Z) + \ldots \nonumber \\
  &+& C_n P_{n}(Z) \Bigr] \; , \; \label{intpol}
\end{eqnarray}
where $P_n(Z)$ are the Legendre polynomials ($P_{0}(Z) = 1$), while $E(1) \approx$ -0.7891967666545 
$a.u.$ and $E(-1) \approx$ -0.51600379066033 $a.u.$ are the total energies of the ground $1^{1}S(L 
= 0)-$state(s) in the ${}^{\infty}$HPs and Ps$_2$ systems, respectively. For this systems we have 
$Z = 1$ and $Z = -1$, respectively. The unknown coefficients $C^{k}_n$ ($k$ = 1, \ldots, $K$) in this 
formula are determined by using the data from direct and accurate numerical computations of the ground 
states in a number of different ${}^{M}h^{+} e^{-}_2 e^{+}$ systems with different `nuclear' masses $M 
(\ge 1)$. Similar formulas can be written for other bound state properties of these four-body systems. 
Numerical values of the first ten coefficients in the formula, Eq.(\ref{intpol}), are presented in 
Table \ref{TablHF}. To determine these ten coefficients we have used data obtained in numerical 
calculations of twenty different positronium hydrides ${}^{M}h^{+} e^{-}_2 e^{+}$. The formula, 
Eq.(\ref{intpol}), with the coefficients from Table \ref{TablHF} provides $\approx$ 8 - 9 correct 
decimal digits for the ground (bound) $1^{1}S(L = 0)-$state energy in an arbitrary exitonic complex 
${}^{M}h^{+} e^{-}_2 e^{+}$ (or positronium hydride), where $M \ge 1$. Ten similar coefficients 
$C^{\star}_{n}$ shown in Table VI have been determined by using computational data for fourteen 
different ${}^{M}h^{+} e^{-}_2 e^{+}$ systems. In reality, the formula, Eq.(\ref{intpol}), with the 
$C^{\star}_{n}$ coefficients provides a better accuracy. Our mass-interpolation formula, 
Eq.(\ref{intpol}), can be used to predict the total energies of a number of `new' ${}^{M}h^{+} e^{-}_2 
e^{+}$ systems. Table VI contains the predicted total energies for the five model ${}^{M}h^{+} e^{-}_2
e^{+}$ systems in which $M = m_{\mu}, 100 \; m_e, 75 \; m_e, 25 \; m_e$ and 15 $m_e$. The overall 
accuracy of our current predictions can be evaluated as $\approx 1 \cdot 10^{-9}$ $a.u.$

Here we have to note that in modern solid state physics there is a strong and permanent interest to 
the bound states in model systems in which masses of the both `holes' and `electrons' can be less 
than unity. Formally, the original idea of Wannier-Mott was based on the assumption that in 
semiconductors with very large electric permittivity $\varepsilon$ (or dielectric tensor) we can 
introduce the two new `effective' particles: (1) negatively charged `electron' which corresponds to 
the bottom of conduction band, and (2) positively charged `hole' which corresponds to the top of 
valence band. The numerical values of masses of these effective particles are determined by assuming 
the exact parabolic symmetry for the minimum of the conduction band and maximum of the valence band. 
There is also a direct Coulomb interaction between these two effective particles. Plus, in many actual 
semi-conductors (e.g., in CdS) we always have a number of different conduction and valence bands 
(usually, two/three bands). This means that in Wannier-Mott theory the masses of `holes' and 
`electrons' are, in fact, some experimental parameters, rather than actual particle masses. In actual 
computations such effective masses of the holes and `electrons' were varied between 0.00985 $m_e$ and 
2.2678 $m_e$ (see, references in \cite{{Bhatia}}, \cite{Dav}, \cite{ShaRod}. Note also that almost all 
numerical computations of few-body exitonic complexes and/or complexes with ionized impurities in 
solid state physics are performed in the quasi-atomic units, which are defined by choosing the energy 
and linear size units as follows 
\begin{eqnarray}
 E_0 = \frac{m e^{4}}{\hbar^{2}} \Bigl(\frac{m^{*}_{e}}{m}\Bigr) \frac{1}{\varepsilon^{2}} \; \; \; 
 {\rm and} \; \; \;  a_0 = \frac{\hbar^{2}}{m e^{2}} \Bigl(\frac{m}{m^{*}_{e}}\Bigr) \varepsilon \; 
 \; . \; \nonumber  
\end{eqnarray}
where $\varepsilon$ is the electric permittivity of semiconductor which is, in fact, a macroscopic 
parameter. For the first time this system of units was used in calculations of the three- and four-body 
exitonic complexes performed in \cite{ShaRod}. 

In this study we also consider the four-body `positronium hydrides' $h^{+} e^{-}_2 e^{+}$ which also contain 
four point particles with unit electrical charges, but the mass of positively charged hole $h^{+}$ is less 
than unity, i.e., less than electron mass in atomic units. Calculations of the bound states in such systems 
is tricky, since all our codes work only for few-body systems where the minimal particle mass is always 
larger (or equal) unity (or $m_e = 1$). Such an agreement has been made at early stages of development of 
few-body codes \cite{FroBi}. However, in few-body systems with unit electrical charges there is the well 
known CMP-invariance, which was discovered in \cite{FroSmi96} for the four-body systems with unit electrical 
charges. In our current four-body case this fact simply means that the energies of the two systems 
${}^{m}h^{+} e^{-}_2 e^{+}$ and $e^{+} Y^{-}_2 Z^{+}$ (expressed in atomic units) are related by the 
following, exact equation: $E({}^{m}h^{+} e^{-}_2 e^{+}) = m E(e^{+} Y^{-}_2 Z^{+})$. Here $m_h = m \le 1$ 
and $M_{Y^{-}} = M_{Z^{+}} = \frac{1}{m} \ge 1$, where all masses are also expressed in atomic units. In 
other words, the total energy of the ${}^{0.1}h^{+} e^{-}_2 e^{+}$ system equals 0.1 of the total energy of 
the $e^{+} Y^{-}_2 Z^{+}$ system, where $M_{Y^{-}} = M_{Z^{+}}$ = 10 $m_e$. For the last $e^{+} Y^{-}_2 
Z^{+}$ system our accurate computational method work perfectly. The same (or very similar) mass-relations 
can also be found for other bound state properties in these `mass-conjugate' four-body systems. This 
CMP-invariance in the four-body systems (in general, in few-body systems) opens a wide avenue for 
applications of our and other computational methods to accurate and highly-accurate calculations of bound 
states in the systems where the minimal mass is smaller than unity. 

The total energies of these four-body systems are shown in Table VI. As follows from the results presented 
in that Table all these model four-body $h^{+} e^{-}_2 e^{+}$ systems are stable against dissociation into 
two neutral (two-body) clusters, i.e., against reaction $h^{+} e^{-}_2 e^{+} = h^{+} e^{-} +  e^{+} e^{-}$ 
which corresponds to the actual (or physical) threshold of stability of the $h^{+} e^{-}_2 e^{+}$ system. 
It is also interesting to note that the actual limit (at $m_{h} \rightarrow 0$) of the total energies of 
four-body $h^{+} e^{-}_2 e^{+}$ systems coincides with the total energy of the three-body Ps$^{-}$ ion ($E 
=$ -0.2620050702329801077704065(55) $a.u.$ \cite{Fro2015}). Another fact can be formulated in the form: our 
mass-interpolation formula, Eq.(\ref{intpol}), is directly applicable to the systems $h^{+} e^{-}_2 e^{+}$ 
with $m_h \le 1$, if we redefine parameters in this formula as follows: $z = 1 - 2 m$, where $m = m_h$ is 
the hole mass expressed in $m_e$. In addition to this, we have to assume that now $E(-1) = E($Ps$_2)$ and 
$E(1) = E($Ps$^{-})$ are the total energies of the ground state(s) in the bi-positronium Ps$_2$ and 
three-body Ps$^{-}$ ion, respectively. After such a re-definition of parameters, the formula, 
Eq.(\ref{intpol}), becomes a good mass-interpolation formula for the total energies of the $h^{+} e^{-}_2 
e^{+}$ systems, where $m_h \le 1$. The explicit form of this mass-interpolation formula in this case is 
\begin{eqnarray}
 E(z) = \Bigl(\frac{1 + z}{2}\Bigr) E(1) + \Bigl(\frac{1 - z}{2}\Bigr) E(-1) + \Bigl(\frac{1 
 - z^{2}}{4}\Bigr) \Bigl[ B_0 &+& B_1 P_{1}(z) + B_2 P_{2}(z) + \ldots \nonumber \\
 &+& B_n P_{n}(z) \Bigr] \; , \; \label{intpolA}
\end{eqnarray}
where $E(1)$ is the total energy of the three-body Ps$^{-}$ ion, while $E(-1)$ is the total 
energy of the four-body Ps$_2$ system. The unknown coefficients $B^{k}_n$ ($k$ = 1, \ldots, 
$K$) in this formula are determined by using the data from direct and accurate numerical 
computations of the ground states in a number of different ${}^{m}h^{+} e^{-}_2 e^{+}$ systems 
with different `nuclear' masses $m (\le 1)$. Analogous mass-interpolation formulas can be 
derived for other bound state properties of these systems, e.g., for the $\langle r_{ij} 
\rangle$ and $\langle \delta_{ij} \rangle$ expectation values. Note also that all Legendre 
polynomials of different orders, which are needed in our calculations, can be determined for 
one numerical value of argument $x$ by using the known recurrence formula (see, e.g., 
\cite{GR}):
\begin{eqnarray}
  P_{n+1}(x) = \Bigl( 2 - \frac{1}{n + 1} \Bigr) x P_{n}(x) - \Bigl( 1 - \frac{1}{n + 1} 
  \Bigr) P_{n-1}(x) 
\end{eqnarray}
and two `initial conditions' $P_{0}(x) = 1$ and $P_{1}(x) = x$. 

Numerical values of the first seven coefficients in the formula, Eq.(\ref{intpolA}), can also 
be found in Table \ref{TablHF}. To determine these ten coefficients we have used data obtained 
in numerical calculations of the ten `model' positronium hydrides ${}^{M}h^{+} e^{-}_2 e^{+}$. 
The formula, Eq.(\ref{intpolA}), with the coefficients from Table \ref{TablHF} provides $\approx$ 
8 - 9 correct decimal digits for the ground (bound) $1^{1}S(L = 0)-$state energy in an arbitrary 
exitonic complex ${}^{m}h^{+} e^{-}_2 e^{+}$ (or positronium hydride), where $m \le 1$. Finally, 
based on the results of calculations for a large number of systems we have found that that our 
mass-interpolation formulas, Eqs.(\ref{intpol}) and (\ref{intpolA}) for the $h^{+} e^{-}_2 e^{+}$ 
systems with $m_h = M \ge 1$ and $m_h = m \le 1$ are reliable and numerically stable. Based on our 
mass-interpolation formula, Eq.(\ref{intpolA}), for the ${}^{m}h^{+} e^{-}_2 e^{+}$ systems we 
have predicted the total energies for five new similar systems where $m = 0.85 \; m_e, 0.75 \; 
m_e, 0.55 \; m_e, 0.3 \; m_e$ and 0.15 $m_e$. The overall accuracy of our current predictions for 
these systems  can be evaluated as $\approx 3 \cdot 10^{-9}$ $a.u.$

Construction of accurate mass-interpolation formulas for the total energies and other bound state 
properties is an absolutely new and advanced area of research in few-body physics. Currently, we 
can see here quite a few directions which must be investigated in the future. For instance, one 
can construct the united mass-interpolation formula for all four-body ions positronium hydrides 
${}^{M}h^{+} e^{-}_2 e^{+}$. Very likely, such a formula cannot be based, in principle, on the 
orthogonal polynomials.  

\section{New methods for highly accurate, bound state computations of the positronium hydrides} 

In our numerical, bound state computations of the four-body positronium hydrides HPs we have used the variational 
expansion written in six-dimensional gaussoids, Eq.(\ref{gauss}), each of which explicitly depends upon the six 
relative coordinates $r_{12}, r_{13}, r_{23}, r_{14}, r_{24}$ and $r_{34}$, where $r_{ij} = \mid {\bf r}_i - {\bf 
r}_j \mid = r_{ji}$ and ${\bf r}_i$ ($i$ = 1, 2, 3, 4) are the Cartesian coordinates of all four (point) particles. 
This variational expansion is well known in few-body physics since the middle of 1970's (see, e.g., \cite{Kol1}, 
\cite{KT}, \cite{Posh} and earlier references therein). After many years of successful calculations of different 
bound states in various nuclear and atomic few-body systems this variational expansion in multi-dimensional 
gaussoids can be  recognized as very effective, fast and sufficiently accurate in 99\% cases, including all 
positronium hydrides HPs discussed in this study. 

However, some important properties of many four-body systems, e.g., all interparticle cusp values, cannot be 
computed directly and accurately with the use of this variational expansion, Eq.(\ref{gauss}). This means we 
do not have any criterion to check the actual accuracy of our two-particle delta-functions. Furthermore, for 
Coulomb systems the variational expansion in multi-dimensional gaussoids, Eq.(\ref{gauss}), always provides 
incorrect asymptotics at large interparticle distances $r_{ij}$. This statement is true for each basis function 
in Eq.(\ref{gauss}) and for any trial wave function constructed with the use of Eq.(\ref{gauss}). As a result 
the actual convergence rate of many bound state properties upon the total number of basis functions $N$ used in 
Eq.(\ref{gauss}) is often non-monotonic. Furthermore, such a rate substantially depends upon algorithms and 
subroutines which have been used for optimization of the non-linear parameters in Eq.(\ref{gauss}). In addition 
to this, description of bound few-body systems in multi-dimensional gaussoids is always more diffuse, than 
similar results obtained with the use of other variational expansions. Briefly, this means that all expectation 
values of interparticle distances $\langle r^{n}_{ij} \rangle$, where $n \ge 1$, obtained by using the 
KT-expansion, Eq.(\ref{gauss}), are systematically and noticeably larger than the same expectation values 
computed by other methods. 

In order to avoid these and other similar problems in accurate, bound state computations of the four-body 
systems in \cite{Our2001} and \cite{Our2003} we have proposed and developed another variational expansion 
which is written in the form of `linear' exponential variational expansion of the four-body relative 
coordinates $r_{ij}$: 
\begin{eqnarray}
 &&\Psi_{N}(r_{14},r_{24},r_{34},r_{12},r_{13},r_{23}) = \; \; \; \nonumber \\ 
 &&{\cal A}_S \sum^{N}_{k=1} C_k \exp(-\alpha^{(k)}_{12} r_{12} -\alpha^{(k)}_{13} r_{13} -\alpha^{(k)}_{14} 
 r_{14} -\alpha^{(k)}_{23} r_{23} -\alpha^{(k)}_{24} r_{24} -\alpha^{(k)}_{34} r_{34}) \; \; \; , \label{exp}
\end{eqnarray}
where $C_k$ are the linear variational parameters of this expansion $(k = 1, \ldots, N$), while $\alpha^{k}_{ij}$ 
are the non-linear parameters. The operator ${\cal A}_S$ is an unitary projector which produces the trial wave 
function of the correct permutation symmetry between all identical particles in the system. The variational 
expansion, Eq.(\ref{exp}), includes the same six interparticle distances which are included in the expansion, 
Eq.(\ref{gauss}). Numerous advantages of this exponential variational expansion, Eq.(\ref{exp}), are quite 
obvious. First of all, the variational expansion, Eq.(\ref{exp}), and each of its basis functions has the correct 
asymptotics at long-range interparticle distances $r_{ij}$. Therefore, this variational expansion, Eq.(\ref{exp}), 
is the truly highly accurate expansion for bound state computations in all four-body systems, where particles 
interact with each other either by the Coulomb, or Yukawa, or exponential potential, or by a potential which is 
represented as a finite linear combination of these potentials. In positronium hydrides the interaction potential 
between each pair of particles is a pure Coulomb potential.   

Second, the actual convergence of the four-body exponential variational expansion, Eq.(\ref{exp}), upon the 
total number of basis functions $N$ used, is significantly faster than for multi-dimensional gaussoids, 
Eq.(\ref{gauss}). In fact, for an arbitrary Coulomb four-body system the exponential expansion, Eq.(\ref{exp}), 
converges faster (and even significantly faster) than other variational expansions currently known and used 
for four-body systems. For many bound state properties the actual convergence of our exponential variational 
four-body expansion, Eq.(\ref{exp}), upon the total number of basis functions $N$ used, is usually monotonic 
and one-sided. Third, accurate numerical calculations of the bound state properties, or expectation values, 
including interparticle cusp values, various delta-functions, etc, are relatively easy to perform by applying 
the exponential expansion, Eq.(\ref{exp}). In fact, for the pure exponential basis set, Eq.(\ref{exp}), one 
finds no difference between three- and four-body cases. Other advantages of our exponential four-body 
variational expansion, Eq.(\ref{exp}), are discussed in \cite{Our2001} and \cite{Our2003}. After these papers 
were published, Frank Harris wrote an effective computer code for this method. By using his code and 40 basis 
functions (four-dimensional exponents) we have found the following total energy of the positronium 
${}^{\infty}$HPs hydride: $E$ = -0.7890967075 $a.u.$ Analogous value of the total energy obtained with the 
use of 50 four-body exponents is $E$ = -0.78913367330 $a.u.$ is very close to the `exact' total energy of 
this system (see above). These results indicate clearly that our exponential variational expansion developed 
in \cite{Our2001} and \cite{Our2003} is highly accurate for bound state computations of arbitrary four-body 
systems. 

Note that the exponential variational expansion, Eq.(\ref{exp}), has nothing to do with the traditional 
atomic Hyllearaas expansion (or James-Coolidge expansion \cite{JC}) used for three-electron atoms and 
ions (see, e.g., \cite{FKing}). Indeed, our exponential variational expansion, Eq.(\ref{exp}), is truly 
correlated four-body expansion which can be applied for highly accurate numerical computations of bound 
states in arbitrary four-body systems (masses of particles can be arbitrary!), and it is not restricted 
to the one-center, three-electron atomic systems only.  

Another interesting four-body variational expansion is written in the form \cite{Fro2010} 
\begin{eqnarray}
 &&\Psi_{N}(r_{14},r_{24},r_{34},r_{12},r_{13},r_{23}) = \; \; \; \nonumber \\ 
 &&{\cal A}_S \sum^{N}_{k=1} C_k r^{n_1(k)}_{23} r^{n_2(k)}_{13} r^{n_3(k)}_{12} \; r^{m_1(k)}_{14} 
 r^{m_2(k)}_{24} r^{m_3(k)}_{34} \;  \exp(-\alpha^{(k)}_{14} r_{14} -\beta^{(k)}_{24} r_{24} 
 -\gamma^{(k)}_{34} r_{34}) \; \; , \; \label{1/2exp}
\end{eqnarray}
where $r_{ij} = r_{ji}$ are the four-body relative coordinates, the operator ${\cal A}_S$ is the 
symmetrization operator, while $\alpha_{i}, \beta_{i}, \gamma_{i}$ are the $3 N$ non-linear parameters 
of this expansion which are varied in actual computations. In Eq.(\ref{1/2exp}) all powers $n_i(k)$ and 
$m_j(k)$ of the relative coordinates are assumed to be non-negative and symbol 4 always designates the 
heaviest particle. The variational expansion, Eq.(\ref{1/2exp}), represents the bound states with $L = 
0$ (or $S-$states, for short). Possible generalizations of this variational expansion to the rotationally 
excited states with $L \ge 1$ are relatively simple and straightforward, but here we cannot discuss that
interesting problem. 

In our earlier paper \cite{Fro2010} the variational expansion, Eq.(\ref{1/2exp}), was called the 
semi-exponential variational expansion in the relative four-body coordinates. In applications to actual 
four-body systems the powers of relative coordinates in Eq.(\ref{1/2exp}) are chosen by the `families' 
(or sub-spaces) of polynomial parts of the basis functions. As is well known (see, e.g., \cite{GelfLA}) 
polynomials of degree $\le \Omega$ form the $(\Omega + 1)-$dimensional vector space. To construct the 
trial functions for the positronium hydride ${}^{\infty}$HPs we have chosen two families of polynomial 
wave functions with $\Omega$ = 28 and $\Omega$ = 84. After a few optimizations our current results 
(total energies) for the ground $S-$state in the ${}^{\infty}$HPs hydride are $E(28) =$ -0.781757154783 
$a.u.$ and $E(84) =$ -0.787961386805 $a.u.$, respectively. These energies are obviously less accurate 
than values obtained above with the use of exponential four-body expansion, Eq.(\ref{exp}), but they 
are much better than than analogous energies obtained with the original James-Coolidge expansion for 
this system (see, e.g., \cite{OurHPs2}): $E(28) =$ -0.778527035 $a.u.$ and $E(84) =$ -0.786802750 
$a.u.$ In other words, our semi-exponential variational expansion, Eq.(\ref{1/2exp}), is significantly 
more flexible and accurate than the standard Hylleraas (or James-Coolidge) expansion developed long 
ago, which is still applied for accurate bound state calculations of many atomic-like (or one-center) 
Coulomb four-body systems. 

To conclude this Section we have to note that two our variational expansions, Eqs.(\ref{exp}) and 
(\ref{1/2exp}), have a great potential for highly accurate, bound state computations in various 
four-body systems. However, at that time these expansions cannot compete with the KT-variational 
expansion in terms of overall accuracy, numerical stability and general convenience in applications. 
It is clear that each of these `new' variational expansions must be substantially improved in order 
to reproduce 7 - 9 stable decimal digits for the ground state energy of all positronium hydrides. 
Then (and only then) these variational expansions can be considered as effective and fast methods  
developed for highly accurate solutions of the Coulomb for-body problem (bound states). The same 
criterion must be applied to other `alternative' methods developed for similar problems, e.g., for 
different molecular orbital methods \cite{ChemRev} and numerous versions of exponentially correlated 
Hylleraas configuration Interactions (E-Hy-CI) \cite{AdvQC}. In any case we did not want to transform 
this short Section into a review of all methods which can be applied, in principle, for highly 
accurate, variational (and non-variational) calculations of the ground state(s) in positronium hydrides. 

\section{Conclusion}

We have investigated the bound state properties of the four-body positronium hydrides ${}^{\infty}$HPs, 
${}^{1}$HPs, DPs, TPs and four-lepton, muonic hydride MuPs. Annihilation of the electron-positron pair(s) 
in all positronuim hydrides is considered in detail. The rates of the total, two- and three-photon 
annihilations of the $(e^{-},e^{+})-$pair in the HPs hydrides have been determined to very good accuracy. 
These  $\Gamma, \Gamma_{2 \gamma}$ and $\Gamma_{3 \gamma}$ rates are the largest values in comparison to 
analogous rates of other annihilation processes in HPs. We have also evaluated a number of other 
annihilation rates, including the $\Gamma_{4 \gamma}, \Gamma_{5 \gamma}$ rates of the four- and 
five-photon annihilations and two different rates of one-photon annihilation $\Gamma^{a}_{1 \gamma}, 
\Gamma^{F}_{1 \gamma}$ of the two different kinds. The hyperfine structure of these posironium hydrides 
is considered. In particular, we have determined the hyperfine structure splitting in each of the 
positronium hydrides ${}^{1}$HPs, DPs, TPs and MuPs. The bound state properties of some four-body exitonic 
${}^{M}h^{+} e^{-}_2 e^{+}$ complexes with unit electrical charges are also briefly discussed. 

We have also investigated the mass-dependence of the total energies and some other bound state properties 
of the four-body exitonic ${}^{M}h^{+} e^{-}_2 e^{+}$ complexes (with $M \ge 1$ and $M \le 1$). Some of 
these systems are similar to the regular positronium hydrides HPs. The both cases, when $M \ge 1$ and $M 
\le 1$ of these exitonic ${}^{M}h^{+} e^{-}_2 e^{+}$ complexes have been considered. Finally, we discuss 
a possibility to apply our new four-body exponential and semi-exponential variational expansions in the 
relative coordinates, Eqs.(\ref{exp}) and (\ref{1/2exp}), for accurate computations of the bound states 
in positronium hydrides. It is shown that the both these variational expansions have a great potential in 
applications to bound state calculations of positronium hydrides and many other four-body systems. The 
both these variational expansions allow one to determine interparticle cusp values and a few other 
properties which cannot be obtained in the direct computations with the use of KT-expansion, 
Eq.(\ref{gauss}). 

\begin{center}
 
   {\bf Acknowledgments}

\end{center}

This paper arose out of a large number of ideas and approaches which I/we have developed 
over many years. 
During this work I have greatly benefited from numerous discussions with 
Anand K. Bhatia, Richard J. 
Drachman, Frank E. Harris, Sergrei I. Kryuchkov, Sergiy Bubin, 
Frederick W. King and others. I would like 
to thank all of them for their valuable help 
and inspiration. Unfortunately, when this paper was in 
preparation, Richard J. Drachman was 
tragically killed in an automobile accident on 6th of April 2021. The 
present contribution 
is dedicated to the memory of this dear friend.

%
\begin{table}[tbp]
\caption{The expectation values $\langle \hat{A} \rangle$ in atomic units $a.u.$ of some bound 
         state properties for the ground (bound) ${}^{1}S(L = 0)-$states of the positronium 
         hydrides ${}^{M}H^{+} e^{-}_2 e^{+}$. In this Table the notations `+' and `-' stand 
         for the positron and electron, respectively, while the symbol `p' means the heavy, 
         positively charged nucleus of the hydrogen isotope.}
     \begin{center}
     \scalebox{0.90}{%
     \begin{tabular}{| l | l | l | l | l | l |}
         \hline\hline
system & ${}^{\infty}$H$^{+} e^{-}_2 e^{+}$ & ${}^{3}H^{+} e^{-}_2 e^{+}$ & ${}^{2}H^{+} e^{-}_2 e^{+}$ & ${}^{1}H^{+} e^{-}_2 e^{+}$ & $\mu^{+} e^{-}_2 e^{+}$ \\ 
       \hline\hline
 $\langle r^{-2}_{--} \rangle$ & 0.213911 & 0.213823 & 0.213779 & 0.213648 & 0.211598 \\

 $\langle r^{-2}_{+-} \rangle$ & 0.349144 & 0.349120 & 0.349120 & 0.349073 & 0.348519 \\

 $\langle r^{-2}_{p-} \rangle$ & 1.207065 & 1.206593 & 1.206358 & 1.205651 & 1.194598 \\

 $\langle r^{-2}_{p+} \rangle$ & 0.172164 & 0.172114 & 0.172089 & 0.172015 & 0.170851 \\
        \hline\hline
 $\langle r^{-1}_{--} \rangle$ & 0.3705554 & 0.3704805 & 0.3704432 & 0.3703313 & 0.3685790 \\

 $\langle r^{-1}_{+-} \rangle$ & 0.4184965 & 0.4184740 & 0.4184628 & 0.4184292 & 0.4179026 \\

 $\langle r^{-1}_{p-} \rangle$ & 0.7297093 & 0.7295588 & 0.7294839 & 0.7292589 & 0.7257325 \\

 $\langle r^{-1}_{p+} \rangle$ & 0.3474623 & 0.3474090 & 0.3473824 & 0.3473027 & 0.3460529 \\
        \hline
 $\langle r_{--} \rangle$ & 3.574770 & 3.575516 & 3.575888 & 3.577003 & 3.594554 \\

 $\langle r_{+-} \rangle$ & 3.480262 & 3.480563 & 3.480713 & 3.481163 & 3.488243 \\

 $\langle r_{p-} \rangle$ & 2.311517 & 2.312063 & 2.312335 & 2.313151 & 2.325995 \\ 

 $\langle r_{p+} \rangle$ & 3.661608 & 3.662232 & 3.662543 & 3.663477 & 3.678178 \\ 
        \hline\hline
 $\langle r^{2}_{--} \rangle$ & 15.87515 & 15.88199 & 15.88539 & 15.89563 & 16.05703 \\ 

 $\langle r^{2}_{+-} \rangle$ & 15.58407 & 15.58716 & 15.58870 & 15.59332 & 15.666248 \\ 

 $\langle r^{2}_{p-} \rangle$ & 7.812899 & 7.816816 & 7.818765 & 7.824627 & 7.917145 \\   

 $\langle r^{2}_{p+} \rangle$ & 16.25421 & 16.26007 & 16.26299 & 16.27176 & 16.41019 \\
               \hline
 $\langle r^{3}_{--} \rangle$ & 84.5457 & 84.6024 & 84.6307 & 84.7156 & 86.0593 \\  

 $\langle r^{3}_{+-} \rangle$ & 84.3699 & 84.3979 & 84.4118 & 84.4537 & 85.1158 \\ 

 $\langle r^{3}_{p-} \rangle$ & 35.2118 & 35.2397 & 35.2535 & 35.2952 & 35.9552 \\ 

 $\langle r^{3}_{p+} \rangle$ & 85.0937 & 85.1425 & 85.1669 & 85.2400 & 86.3974 \\ 
               \hline
 $\langle r^{4}_{--} \rangle$ & 527.879 & 528.372 & 528.617 & 529.354 & 541.059 \\ 
 
 $\langle r^{4}_{+-} \rangle$ & 532.936 & 533.195 & 533.323 & 533.711 & 539.848 \\ 

 $\langle r^{4}_{p-} \rangle$ & 198.871 & 199.089 & 199.198 & 199.525 & 204.726 \\ 

 $\langle r^{4}_{p+} \rangle$ & 516.036 & 516.456 & 516.665 & 517.294 & 527.286 \\  
        \hline\hline
 $\langle -\frac12 \nabla^{2}_{-} \rangle$ & 0.32617335 & 0.32606754 & 0.32601491 & 0.32585679 & 0.32338494 \\

 $\langle -\frac12 \nabla^{2}_{+} \rangle$ & 0.13685044 & 0.13684430 & 0.13684125 & 0.13683208 & 0.13668953 \\
        \hline\hline
  \end{tabular}}
  \end{center}
  \end{table}
\begin{table}[tbp]
\caption{The total energies $E$ and expectation values of some interparticle delta-functions $\langle \delta_{ij} 
         \rangle$ in atomic units $a.u.$ for the ground (bound) ${}^{1}S(L = 0)-$states of the positronium hydrides 
         ${}^{M}H^{+} e^{-}_2 e^{+}$. In this Table the notations `+' and `-' stand for the positron and electron, 
         respectively, while the symbol `p' means the heavy, positively charged nucleus of the hydrogen isotope.}
     \begin{center}
     \scalebox{0.95}{%
     \begin{tabular}{| l | l | l | l | l | l |}
         \hline\hline
system & ${}^{\infty}H^{+} e^{-}_2 e^{+}$ & ${}^{3}H^{+} e^{-}_2 e^{+}$ & ${}^{2}H^{+} e^{-}_2 e^{+}$ & ${}^{1}H^{+} e^{-}_2 e^{+}$ & $\mu^{+} e^{-}_2 e^{+}$ \\ 
       \hline\hline
 $E$                 & -0.7891967651 & -0.789087847 & -0.789033651 & -0.788870705 & -0.786317295 \\
        \hline
 $\langle \delta_{+-} \rangle$ & 2.448351$\cdot 10^{-2}$ & 2.448163$\cdot 10^{-2}$ & 2.448070$\cdot 10^{-2}$ & 2.447789$\cdot 10^{-2}$ & 2.443306$\cdot 10^{-2}$ \\ 

 $\langle \delta_{p+} \rangle$ & 1.625353$\cdot 10^{-3}$ & 1.624877$\cdot 10^{-3}$ & 1.624642$\cdot 10^{-3}$ & 1.623930$\cdot 10^{-3}$ & 1.612725$\cdot 10^{-3}$ \\ 

 $\langle \delta_{+--} \rangle$ & 3.69845$\cdot 10^{-4}$ & 3.69619$\cdot 10^{-4}$  & 3.69506$\cdot 10^{-4}$  & 3.69168$\cdot 10^{-4}$  & 3.63833$\cdot 10^{-4}$ \\ 
        \hline\hline
  \end{tabular}}
  \end{center}
  \end{table}
\begin{table}[tbp]
\caption{A number of the few-photon annihilation rates $\Gamma_{n \gamma}$ (in $sec^{-1}$) predicted for the ground (bound) 
         $1^{1}S_{e}(L = 0)-$states in the positronium hydrides MuPs, ${}^{1}$HPs, DPs, TPs and ${}^{\infty}$HPs.} 
\label{TablAnhl}
     \begin{center}
     \scalebox{1.05}{%
     \begin{tabular}{| c | c | c | c | c | c |}
        \hline\hline
     rate & ${}^{\infty}$HPs & TPs & DPs & ${}^{1}$HPs & MuPs \\  
        \hline
 $\Gamma$            & 2.5237439$\cdot 10^{9}$ & 2.5235504$\cdot 10^{9}$ & 2.5234541$\cdot 10^{9}$ & 2.5231646$\cdot 10^{9}$ & 2.5185995$\cdot 10^{9}$ \\ 

 $\Gamma_{2 \gamma}$ & 2.4568126$\cdot 10^{9}$ & 2.4566243$\cdot 10^{9}$ & 2.4565305$\cdot 10^{9}$ & 2.4562487$\cdot 10^{9}$ & 2.4518047$\cdot 10^{9}$ \\ 

 $\Gamma_{3 \gamma}$ & 6.6931229$\cdot 10^{7}$ & 6.6926097$\cdot 10^{7}$ & 6.6923543$\cdot 10^{7}$ & 6.6915865$\cdot 10^{7}$ & 6.6794796$\cdot 10^{7}$ \\ 

 $\Gamma_{4 \gamma}$ & 3.6320642$\cdot 10^{3}$ & 3.6317857$\cdot 10^{3}$ & 3.6316471$\cdot 10^{3}$ & 3.6312305$\cdot 10^{3}$ & 3.6246606$\cdot 10^{3}$ \\ 

 $\Gamma_{5 \gamma}$ & 63.91944 & 63.91454 & 63.91210 & 63.90477 & 63.78915 \\ 
       \hline
 $\Gamma^{a}_{1 \gamma}$ & 0.39417 & 0.39392 & 0.39380 & 0.39344 & 0.38776 \\  

 $\Gamma^{F}_{1 \gamma}$ & 1.48715 & 1.48704 & 1.48698 & 1.48681 & 1.48412 \\ 
    \hline\hline 
  \end{tabular}}
  \end{center}
  \end{table}
\begin{table}[tbp]
\caption{The hyperfine structure splittings $\Delta E_{hyps}$ (in $MHz$) for the ground (bound) $1^{1}S_{e}(L = 0)-$states 
         in the positronium hydrides MuPs, ${}^{1}$HPs, DPs and TPs} 
\label{TablHF}
     \begin{center}
     \scalebox{1.05}{%
     \begin{tabular}{| c | c | c | c | c |}
        \hline\hline
 system & factor ${\cal A}$ & $I_{H}$ & $\langle \delta_{p +} \rangle^{(a)}$ & $\Delta E_{hyps}$ \\  
           \hline\hline 
       MuPs  & 14229.176639 & $\frac12$ & 1.621275$\cdot 10^{-3}$ & 23.0693728 \\

 ${}^{1}$HPs & -4469.881841 & $\frac12$ & 1.623930$\cdot 10^{-3}$ & -7.2519050 \\

         DPs &  -686.154150 &     1     & 1.624642$\cdot 10^{-3}$ & -1.6721323 \\
 
         TPs & -4767.754403 & $\frac12$ & 1.624877$\cdot 10^{-3}$ & -7.2630081 \\
    \hline\hline 
  \end{tabular}}
  \end{center}
  ${}^{(a)}$The expectation value of the positron-nucleus delta-function in atomic units.
  \end{table}
\begin{table}[tbp]
\caption{The expectation values $\langle \hat{A} \rangle$ in atomic units $a.u.$) of some bound state properties 
         for the ground (bound) ${}^{1}S(L = 0)-$states of the model positronium hydrides ${}^{M}h^{+} e^{-}_2 
         e^{+}$. In this Table the notations `+' and `-' stand for the positron and electron, respectively, while 
         the symbol `h' means the heavy, positively charged quasi-nucleus with mass $M (= M m_e)^{(a)}$.}
\label{HmPs}
     \begin{center}
     \scalebox{1.05}{%
     \begin{tabular}{| l | l | l | l | l | l |}
        \hline\hline
 system & $E$ & $\langle r_{+-} \rangle$ & $\langle r_{--} \rangle$ & $\langle r_{h -} \rangle$ & $\langle \delta({\bf r}_{+-}) \rangle$ \\  
        \hline\hline
  ${}^{12}h^{+} e^{-}_2 e^{+}$ & -0.744058235415 & 3.612337700 & 3.940177538 & 2.553742718 & 0.02394728 \\ 
 
  ${}^{11}h^{+} e^{-}_2 e^{+}$ & -0.740373921176 & 3.623763252 & 3.964626240 & 2.574977893 & 0.02390840 \\  
        \hline\hline 
 ${}^{10}h^{+} e^{-}_2 e^{+}$ & -0.736033440341 & 3.637343176 & 3.993755510 & 2.600284075 & 0.02385773 \\

 ${}^{9}h^{+} e^{-}_2 e^{+}$ & -0.730843941385 & 3.653774543 & 4.029094985 & 2.630980145 & 0.02380552 \\

 ${}^{8}h^{+} e^{-}_2 e^{+}$ & -0.724528577928 & 3.674032011 & 4.072811334 & 2.668967075 & 0.02373458 \\ 

 ${}^{7}h^{+} e^{-}_2 e^{+}$ & -0.716675039870 & 3.699651073 & 4.128317923 & 2.717213004 & 0.02366023 \\

 ${}^{6}h^{+} e^{-}_2 e^{+}$ & -0.706640804107 & 3.733036217 & 4.201037748 & 2.780491629 & 0.02355747 \\
    \hline\hline 
 ${}^{5}h^{+} e^{-}_2 e^{+}$ & -0.693365530158 & 3.778339650 & 4.300413160 & 2.867138301 & 0.02343085 \\
 
 ${}^{4}h^{+} e^{-}_2 e^{+}$ & -0.674963081481 & 3.843250936 & 4.444240116 & 2.993047371 & 0.02326605 \\

 ${}^{3}h^{+} e^{-}_2 e^{+}$ & -0.647714874405 & 3.943731008 & 4.670529754 & 3.192898318 & 0.02302489 \\

 ${}^{2}h^{+} e^{-}_2 e^{+}$ & -0.603091391970 & 4.118241595 & 5.076718755 & 3.559953733 & 0.02266962 \\
    \hline\hline 
  \end{tabular}}
  \end{center}
  ${}^{(a)}$The total energy of the four-body bi-positronium molecule Ps$_2$, or ${}^{1}h^{+} e^{-}_2 e^{+}$ 
  positronium hydride in our current notations, is -0.51600379066033 $a.u.$
  \end{table}
\begin{table}[tbp]
\caption{The coefficients $C_{k}$ in the ten-term mass-interpolation formula, Eq.(\ref{intpol}) for the total 
         energies of the ground (bound) $1^{1}S_{e}(L = 0)-$states in the four-body ${}^{M}h^{+} e^{-}_2 e^{+}$ 
         systems, where $M \ge 1$. The coefficients $B_k$ in the seven-term mass-interpolation formula, 
         Eq.(\ref{intpolA}) for the total energies of the ground (bound) $1^{1}S_{e}(L = 0)-$states in the 
         four-body ${}^{m}h^{+} e^{-}_2 e^{+}$ systems, where $m \le 1$.} 
\label{TablHF}
     \begin{center}
     \scalebox{1.05}{%
     \begin{tabular}{| c | c | c | c | c | c |}
        \hline\hline
     $k$ & $C_{k}$ & $C^{\star}_{k}$ & $B_{k}$  & system & $E{}^{(a)}$ (predicted) \\  
           \hline\hline 
 1 &  0.1760927482 &  0.2084671422 & -0.1734561261 & $\mu^{+} e^{-}_2 e^{+}$ & -0.7863172536 \\

 2 &  0.1699737489 &  0.0861748927 & -0.0601985845 & ${}^{100}h^{+} e^{-}_2 e^{+}$ & -0.7832819721 \\

 3 & -0.0831406229 &  0.0204254266 & -0.0132568021 & ${}^{75}h^{+} e^{-}_2 e^{+}$ & -0.7813436733 \\

 4 &  0.1018702975 &  0.0102296923 & -0.0029164214 & ${}^{25}h^{+} e^{-}_2 e^{+}$ & -0.7664017141 \\ 

 5 & -0.0642471972 & -0.0016461962 &  0.0006601439 & ${}^{15}h^{+} e^{-}_2 e^{+}$ & -0.7523881511 \\ 

 6 &  0.0365809780 &  0.0031380338 &  0.0004238099 & ${}^{0.85}h^{+} e^{-}_2 e^{+}$ & -0.4951481654 \\

 7 & -0.0152567422 & -0.0015202684 &  0.0010964638 & ${}^{0.75}h^{+} e^{-}_2 e^{+}$ & -0.4792885124 \\ 

 8 &  0.0048478700 &  0.0007979493 & ------------- & ${}^{0.55}h^{+} e^{-}_2 e^{+}$ & -0.4416910665 \\ 

 9 & -0.0011064536 & -0.0002802924 & -------------- & ${}^{0.3}h^{+} e^{-}_2 e^{+}$ & -0.3786180518 \\ 
 
 10 & 0.0002520781 &  0.0001083263 & -------------- & ${}^{0.15}h^{+} e^{-}_2 e^{+}$ & -0.3279908164$^{(a)}$ \\ 
    \hline\hline 
  \end{tabular}}
  \end{center}  
   ${}^{(a)}$Also, the total energy of the four-body ${}^{0.005}h^{+} e^{-}_2 e^{+}$ system is 
   -0.26450300278 $a.u.$, which is close to the total energy of the three-body Ps$^{-}$ ion ($E \approx$ 
-0.2620050702329801$\ldots$ $a.u.$, see, the main text). 
   \end{table} 

\end{document}